\documentclass[11pt,letter]{article}

\usepackage{amsmath, amsfonts, amssymb, epsfig, pstricks,pst-node,pst-tree,graphicx,./mylay}
\usepackage{flushend}
\usepackage{color}

\usepackage{graphics}
\usepackage{latexsym}
\usepackage{pifont}
\usepackage{balance}
\usepackage{flushend}
\usepackage{xspace}
\usepackage{tabularx}
\usepackage{multirow}

\newcommand{\Ml}{M_{{}_L}}
\newcommand{\Mlpl}{M_{{}_{L+\ell}}}
\newcommand{\Mr}{M_{{}_{r\ell}}}

\newcommand{\Nl}{{\cal T}_{{}_L}}

\newcommand{\Nlmr}{{\cal T}_{{}_{L-r\ell}}}
\newcommand{\nm}{\Nl}
\newcommand{\pl}{{\bf p}_{{}_L}}

\newcommand{\plpl}{{\bf p}_{{}_{L+\ell}}}
\newcommand{\dist}{D}
\newcommand{\rd}{r(\dist)}
\newcommand{\chn}{\omega}
\newcommand{\inpd}{P}
\newcommand{\optd}{Q}
\newcommand{\ptype}{p}
\newcommand{\qtype}{q}

\newcommand{\prob}{\mathbb P}

\pssetlength{\unitlength}{1mm}
\pssetlength{\linewidth}{5pt}

\renewcommand{\c}{c}
\newcommand{\iid}{\emph{iid}\xspace}
\newcommand{\epoch}{t}
\newcommand{\type}{\tau}
\newcommand{\im}{I_m}
\newcommand{\x}{{\bf x}}
\newcommand{\rY}{{\bf Y}}

\newcommand{\X}{{\bf X}}

\newcommand{\y}{{\bf y}}
\newcommand{\B}{\tilde{B}}
\newcommand{\poly}{{\textrm{poly}}}

\newcommand{\beq}{\begin{equation}}
\newcommand{\eeq}{\end{equation}}

\newcommand{\E}{\mathbb E}
\title{Optimal Lempel-Ziv based lossy compression for memoryless data: how to make the right mistakes} 
\author{ 
Narayana P. Santhanam\\
Dept. of Electrical Engg,\\ 
University of Hawaii at Manoa\\ 
\texttt{nsanthan@hawaii.edu}
\and
Dharmendra Modha\\
IBM Almaden Research Center\\ 
\texttt{dmodha@almaden.ibm.com}}
\begin{document}
\maketitle

\begin{abstract}
  \textbf{ Please note: this document is very much work in progress till 10/31 by my estimate. If something seems off,
    it probably is. Please email N Santhanam above---you can get notes for clarification.}  Compression refers to
  encoding data using bits, so that the representation uses as few bits as possible. Compression could be lossless:
  \emph{i.e.}  encoded data can be recovered exactly from its representation) or lossy where the data is compressed more
  than the lossless case, but can still be recovered to within prespecified distortion metric. In this paper, we prove
  the optimality of \emph{Codelet Parsing}, a quasi-linear time algorithm for lossy compression of sequences of bits
  that are independently and identically distributed (\iid) and Hamming distortion. Codelet Parsing extends the lossless
  Lempel Ziv algorithm to the lossy case---a task that has been a focus of the source coding literature for better part
  of two decades now.

  Given \iid sequences $\x$, the expected length of the shortest lossy representation such that $\x$ can be
  reconstructed to within distortion $\dist$ is given by the rate distortion function, $\rd$.  We prove the optimality
  of the Codelet Parsing algorithm for lossy compression of memoryless bit sequences.  It splits the input sequence
  naturally into phrases, representing each phrase by a \emph{codelet}, a potentially distorted phrase of the same
  length.  The codelets in the lossy representation of a length-$n$ string ${\x}$ have length roughly $(\log n)/\rd$,
  and like the lossless Lempel Ziv algorithm, Codelet Parsing constructs codebooks logarithmic in the sequence length.

\end{abstract}

\setcounter{section}{0}

\section*{Introduction}
Kac's lemma~\cite{Kac47} for stationary ergodic sources formalizes the connection of the \emph{recurrence} time of
events with their probabilities. This connection implies an elegant way to recursively compress sequences from
stationary ergodic sources to their entropy, formalized by the Lempel Ziv algorithm for lossless compression.

The theoretical and commercial importance of the Lempel Ziv algorithm and its variants have not only been established
for compression problems, but also for classification~\cite{Ziv08} and denoising~\cite{WOSVW05} algorithms. In addition
to their theoretical guarantees, these algorithms have attractive computational and storage properties, are often
entirely data driven, and do not rest on sensitive choices of parameter values. It is thus not surprising that Lempel
Ziv based algorithms form the core of compression algorithm software, including WINZIP, \verb+gzip+, and the UNIX
\verb+compress+ algorithms.  Additionally, Lempel Ziv compression has had profound influence in the study of complexity,
see for example,~\cite{LZ76,KS87}. For many researchers, this angle perhaps outweighs even the commercial significance
of Lempel Ziv compressors.

\subsection*{Lossy compression} 
Surprisingly, no algorithms as attractive and simple as the Lempel Ziv algorithm are known for lossy compression. In
fact, in the recent past, some researchers were pessimistic about the problem in general, see~\cite{SM11:all} for
details. For example, ~\cite[p.  2709]{berger:gibson} noted that ``All universal lossy coding schemes found to date lack
the relative simplicity that imbues Lempel-Ziv coders and arithmetic coders with economic viability''.

Of course, a lot of research continues on lossy source compression algorithms, mainly with an eye on the potential
theoretical and practical benefits of having such algorithms.

\subsection*{Prior work}
\label{sec:work} We present a representative, but necessarily brief and non-exhaustive review of various known lossy
coding schemes, focussing on algorithmic results. For references to earlier results on existence of universal lossy
codes involving exponential-time constructions, see, Kieffer~\cite{kieffer:93}.  We confine our discussion here to
finite discrete source and reproduction alphabets; for an extensive survey of results for real-valued sources,
see~\cite{gray:neuhoff}.  Among these, we are particularly interested in papers that have focussed on lossy extensions
of the Lempel-Ziv algorithm.

Most algorithms have naturally used approximate string matching~\cite{atallah:01,navarro} instead of exact string
matching as in the Lempel-Ziv algorithms.  The unresolved question has always been which of the ``approximately
matching'' representations to choose. Cheung and Wei~\cite{cheung:wei} extended a \emph{move-to-front} algorithm to
lossy source coding. The algorithm is sup-optimal~\cite{yang:kieffer:98}. Later, Zhang and Wei~\cite{zhang:wei:96}
proposed an universal, on-line lossy coding algorithm for the fixed-rate case. Morita and
Kobayashi~\cite{morita:kobayashi} extended the LZW algorithm, but their algorithm is known to be sub-optimal for
memoryless sources~\cite{yang:kieffer:98}. Constantinescu and Storer~\cite{cornel:storer:93,cornel:94} combined ideas
from lossless Lempel-Ziv algorithms and vector quantization to design first \emph{practical} implementations of lossy
image compression based on approximate string matching.  The problem of ``selecting amongst multiple matches'' mentioned
above was termed the ``Match Heuristic'' in their work; see, also, Storer~\cite[p.  111]{storer:88}.  Steinberg and
Gutman~\cite{steinberg:gutman} and Luczak and Szpankowski~\cite{luczak:szpankowski} considered the fixed-database
version of the Lempel-Ziv algorithm, and provided sub-optimal performance guarantees. However, Yang and
Kieffer~\cite{yang:kieffer:98} established that all previous fixed-database extensions of the Lempel-Ziv algorithm are
suboptimal.

Kontoyiannis~\cite{kontoyiannis} presented a scheme where multiple databases are used at the encoder, which must also be
known to the decoder. However, when the reproduction alphabet is large, the number of training databases is unreasonably
large. Atallah et al.~\cite{genin} considered a cubic-time, adaptive algorithm (PMIC) in the spirit of LZ77. Their
algorithm is not sequential in the sense of~\cite{kieffer:yang:seq}, since its encoding delay grows faster than
$o(n)$. Alzina et al.~\cite{alzina} combined ideas from~\cite{genin} and~\cite{cornel:storer:93,cornel:94} to propose a
2D-PMIC algorithm that is more suited for two dimensional images.

Continuing the quest for Lempel-Ziv-type lossy algorithms, Zamir and Rose~\cite{zamir:rose:haifa} further studied the
algorithm in~\cite{morita:kobayashi}. From the multiple codewords that may match a source word, they suggest choosing
one ``at random''. From a theoretical perspective, by assuming uniqueness, Zamir and Rose~\cite{zamir:rose} proposed a
natural type selection scheme for finding the type of the optimal reproduction distribution. In later work, Kochman and
Zamir~\cite{kochman:zamir} pointed out that the theoretical procedure in~\cite{zamir:rose} is in itself not practical
and demonstrated an application of natural-type selection to on-line codebook selection from a parametric class.  Along
a different line, Yang and Kieffer~\cite{yang:kieffer:96} have proposed exponential-time Lempel-Ziv-type block codes
that are universal (for stationary, ergodic sources and for individual sequences). In a related work, Yang and
Zhang~\cite{yang:zhang:mar99} presented fixed-slope universal lossy coding schemes that search for the reproduction
sequence through a trellis in a fashion reminiscent of the Viterbi algorithm.

The lossy coding problem has been approached using methods fundamentally different from the Lempel Ziv like approaches
as well. Matsunaga and Yamamoto~\cite{MatYam03} considered LDPC codes for lossy data compression. In this line of work,
Wainwright and Maneva~\cite{WaiMan05} looked at message passing and Low Density Generator matrices (LDGM), while
Martinian and Wainwright~\cite{MarWai06a} looked into the construction of LDGMs and compound code constructions, showing
the existence of compound LDGM-LDPC constructions that achieve the rate-distortion bound.  Futher bounds on the
performance of these constructions have been considered in~\cite{DimWaiRam07}. In another line of attack, Jalali,
Montanari and Weissman approach the problem using dynamic programming approaches~\cite{JMW10}.

\subsection*{Challenges}

In this paper, we consider lossy encoding of memoryless data. What constitutes progress at a conceptual level? The
algorithm we consider, Codelet Parsing, reduces to the Lempel Ziv algorithm (LZ78 version) for lossless encoding, and we
believe that Codelet Parsing may be optimal for stationary ergodic sources as well.

One way to think of lossy encoding is as follows. We construct a \emph{codebook} $\cC$, a set of sequences substantially
smaller than the set of all possible sequences. Given any sequence $\x$, we fix an element of $\cC$ as its
representation. Thus, for any sequence $\x$, we only have to describe which element in $\cC$ it maps to (rather than all
possible sequences). If $\cC$ has been chosen well, every sequence has some sequence of $\cC$ that is fairly close to
it. Thus the crux of the lossy compression problem is (i) to construct $\cC$, and (ii) to search for a
representation. The minimum size of $\cC$ is characterized through the rate distortion function $\rd$.

We sketch a rough picture of the problem of lossy compression now.  While not necessary for the results of our paper,
most of the statements below can be made formal. If a length-$n$ sequence $\X$ is generated \iid Bernoulli $p$, the
probability $\X$ matches a length-$n$ sequence $\y$ to within distortion $\dist$ is highest if the type of $\y$ is
$(p-\dist)/(1-2\dist)$. The probability of match is then $2^{-n\rd}/\poly(n)$. Thus if we are to encode length $n$
sequences, $|\cC|\ge\poly(n)2^{n\rd}$ in order to satisfy the distortion budget $\dist$. In fact, a randomly chosen
$\cC$ from sequences with type $(p-\dist)/(1-2\dist)$ will cover almost all input sequences with size
$|\cC|=\poly(n)2^{n\rd}$. Thus \emph{random coding} uses $\ge n\rd+\cO\Paren{\log n}$ bits to represent a string. This
approach is clearly not practical (both construction and search take exponential time) and we look for more efficient
ways to achieve the goal by using more structured codebooks.

Lempel Ziv approaches circumvent the problem of exponential encoding and search time with a recursive
construction. Rather than construct codebooks for length $n$ sequences, one constructs a set $\cD$ of sequences of
length $\frac{(\log n)}{\rd}$. Often, codebooks over lengths smaller than the sequence length are refered to as
\emph{dictionaries} in Lempel-Ziv literature to avoid confusion, and we adopt the same convention. The algorithm splits
the length-$n$ sequence $\X$ into \emph{phrases} of length $\frac{(\log n)}{\rd}$, representing each phrase by one of
the elements of $\cD$. The strength of this approach is that the construction of $\cD$ happens naturally using just the
data to be encoded, and is known to capture the probability laws governing the data as long as the data is stationary
ergodic (not just memoryless).

Furthermore, a simple argument about recurrence time of events shows that it is not possible to estimate probabilities
of all strings of length $\Omega(\log n)$ using $n$ samples---a fact that will come into play if the algorithms are to
be extended for all stationary ergodic sources. Thus, the dictionaries cannot be over sequences longer than $\cO(\log
n)$ if we have the goal of extending our algorithm to all stationary ergodic sources.

What should we expect from all this? We should expect an approach using the Lempel Ziv theme to have redundancy (the
excess bits over the rate distortion $n\rd$ term) commensurate with random encoding of sequences of length $(\log
n)/\rd$.  Comparing with the numbers given above for random encoding of length $n$ sequences, we conclude that such
approaches use $n\rd+\cO(\frac{n\log\log n}{\log n})$ for length $n$ sequences.  However, the complexity of search
through $\cD$ to represent any phrase of length $(\log n)/\rd$ is linear in $n$, leading to an overall complexity of
$\cO(n^2/(\log n))$ in order to encode a sequence of length $n$.

Note that actually adapting the Lempel Ziv theme is non-trivial. In particular, how does one guarantee that the
dictionary $\cD$ constructed does match the performance of a randomly chosen and \emph{good} codebook of length $(\log
n)/\rd$?  This is analogous to the channel coding problem for communication, where a randomly chosen code is good with
high probability---yet constructing practical codes that are optimal took almost 60 years of intense research. Indeed,
the connections run deeper---lossy compression is a \emph{covering} problem, while channel coding is a \emph{packing}
problem.

Here we show that Codelet Parsing built on the Lempel Ziv theme has a redundancy of $\cO(\frac{\log\log n}{\log n})$ as
expected. However, Codelet Parsing constructs the dictionary $\cD$ in a more structured manner than brute force random
construction, and finding a match requires only $\poly(\log n)$ (not linear) complexity on an average. Thus Codelet
Parsing is a quasi \emph{linear} algorithm. At the level of encoding length-$n$ sequences, this is seemingly only an
improvement from quadratic to linear complexity (notwithstanding the fact that it is not even clear how to achieve
quadratic complexity), but such an improvement also indicates a new way to build the dictionary.

\subsection*{Contributions}
This paper builds on the Lempel Ziv approach along the lines of~\cite{modha:dcc:03,modha:lossy:RJ2, SM11:all}. In
particular, we analyze an idealization of a Lempel Ziv like algorithm called Codelet Parsing, proposed by the authors
in~\cite{MS05}.  In a preliminary paper~\cite{SM11:all}, we showed convergence of Codelet Parsing's coding rate (the
number of bits used to describe the lossy representation of a string, normalized by the length of the string) to the
rate distortion function, when the input string is \iid and the distortion is fixed to be Hamming distortion.

In this paper, we obtain a covering lemma that allows us to characterize the rate of convergence of the coding rate as
$\cO\Paren{\frac{\log\log n}{\log n}}$ (exponentially better than the loose estimate in~\cite{SM11:all}). It is
important to highlight how this result substantially strengthens~\cite{SM11:all}.

In particular, we note a few important points. The distorted phrases are of length roughly $(\log n)/\rd$ and are
obtained by searching through a codebook (maintained as a complete binary tree as in the LZ78 setup).
\begin{enumerate}
\item The sequences in this codebook are not obtained by exhaustive search. Instead, they are recursively obtained by
  calling on codebook constructions over shorter lengths of length $\cO(\log\log n)$. In addition, searching for an
  approximate match does not require an exhaustive search over sequences of length $(\log n)/\rd$.
\item The shorter codebook constructions work in synergy in a manner of speaking since convergence to $\rd$ is
  $\cO\Paren{\frac{\log\log n}{\log n}}$. This rate is almost what we should expect even for exhaustive codebook
  constructions of length $\cO(\log n)$.
\end{enumerate}
A consequence of the first point is that we obtain an algorithm that is quasi-\emph{linear} (linear with log factors)
complexity. This is a savings from the potentially super-quadratic complexity if we exhaustively construct or search
through codebooks of length $\log n/\rd$,

To put the second point in perspective, the convergence rate of our algorithm is exponentially faster than what could
have been obtained by partitioning ${\x}$ into phrases of length $\cO(\log\log n)$, and representing each phrase in a
lossy manner using a codebook of length $\cO(\log \log n)$.

\section{Preliminaries and combinatorial interpretations}

\subsection{Rate-Distortion and Lower-Mutual-Information}

Let $X^n=X_1,X_2,\ldots$, where $X_i\in\sets{0,1}$ for all $i$, be a realization of an \iid process $\inpd$, with the
marginal distribution on $X_i$ being $\inpd(X_i=1)=\ptype$. We represent a string of length $n$, $X^n$ using a
potentially distorted $Y^n\in\sets{0,1}^n$. Let $d(X^n,Y^n)$ denote the Hamming distortion between $X^n$ and $Y^n$. We
adhere to an expected distortion constraint, namely $Ed(X^n,Y^n)<\dist$.  It is customary to call $Y^n$ the
\emph{codeword} used for the lossy representation of $X^n$.  Note that $Y^n$ is not necessarily \iid and is determined
by the algorithm used to pick codewords.

The \emph{rate distortion} function captures, asymptotically, the minimum number of bits that have to be used to
describe strings of length $n$ to within distortion $\dist$. Interestingly, it has a \emph{single letter}
characterization, meaning that it can be specified by looking at the joint distribution over a pair of bits $(Y,X)$ such
that $P(X=1)= p$. The conditional distributions on $X$ given $Y$ correspond to a \emph{channel}, while $Y$ is
interpreted as the channel input and $X$ the channel output.

Let $\cal W$ be the set of all possible channels. The
\emph{rate-distortion} function is 
\[
\rd=R(\inpd, \dist) = 
\min_{
\substack{ 
q',\chn \in {\cal W}: Y\sim q', X\sim p\\
\E d(X,Y) \leq \dist}} 
I (X,Y)
\]
where $I(X,Y)$ denotes the {\em mutual information} and $Y\sim q'$ means $P(Y=1)=q'$.

The lossy coding problem is essentially a covering problem. Suppose we consider length-$n$ sequences $X^n$ generated by
an \iid measure $\inpd$, satisfying $P(X_i=1)=p$ (as befor).  Say we want the the probability of length $n$ sequences
of type $\ptype$ that are within distortion $\dist$ from a sequence $\vecy$ with type $\qtype$.  This probability again
has a single letter characterization in terms of a pair of binary variables $(Y,X)$, where $Y\sim q$ and $X\sim p$.  In
particular, we define
\[
\im(\qtype,\ptype,\dist) 
\ed 
\min_{ 
\substack{ 
\chn\in{\cal W}: X\sim p Y\sim q\\
d(\qtype,\chn)\le \dist }}
I(X,Y),
\]
where we are minimizing the mutual information $I(X,Y)$ over all joint distributions consistent with the marginals being
$X\sim p$ and $Y\sim q$, and $\E d(X,Y)\le \dist$. The probability we want is then $2^{-n
  I_m(\qtype,\ptype,\dist)+\cO(\log n)}$.  $\im(\qtype,\ptype,\dist)$ is a convex function of $\qtype$ for a fixed
$\ptype$, with a minimum at the \emph{optimal reproduction type} $\qtype^*$.

Intuitively speaking, codewords with the optimal reproduction type have the largest $\dist-$balls among sequences of
type $\inpd$, hence, yield the best covering. For a precise formulation of the above concepts,
see~\cite{zhang:wei:96,zamir:rose}. However, to just obtain the estimates given above, a simple combinatorial
calculation followed by picking the dominant term suffices.

%\begin{figure}[!tb]
%\centerline{ \includegraphics[height=1.7in]{Implot.pdf} }
%\centerline{ \includegraphics[height=1.7in]{Implot.jpg}}
%\label{fig:0} \caption{\small An example of lower mutual information.}
%\end{figure}

\subsection{Ballot box problem}
We have an expected distortion constraint between a sequence $X^n$ generated by $\inpd$ and its codeword $Y^n$. As we
will will see, we obtain $Y^n$ by first breaking $X^n$ into disjoint phrases $X^n=\X^{(1)}\upto \X^{(r)}$ (where
$r=\cO(n/\log n)$), and representing each phrase $\X^{(i)}$ by a \emph{codelet} $\y^{(i)}$ of the same length, such that
$d(\x^{(i)},\y^{(i)})\le D$. Such an approach however leads to lack of sufficient structure in the codebooks generated,
leading to quadratic complexity for the algorithm.

To better implement search and representation among codelets, we impose a more restrictive constraint in picking
codelets. We will require not only that $d(\x^{(i)},\y^{(i)})\le D$ in the example above, but that every prefix of
$\x^{(i)}$ be within distortion $D$ of the corresponding prefix of $\y^{(i)}$. Namely, for any $l$, if $\x'$ and $\y'$
are $l$-length prefixes of $\x^{(i)}$ and $\y^{(i)}$ respectively, we require that $d(\x',\y')\le D$ as well. We then
write $\x^{(i)}\sim \y^{(i)}$ and say that $\x^{(i)}$ \emph{matches} $\y^{(i)}$.

The important thing is that the probability that a codelet $\y$ finds a match is essentially the probability of all
sequences with distortion $\dist$ from $\y$. In fact

(maybe state stronger too?)  
\bLemma Let length $n$ sequences $\X$ be generated by an \iid source $P$, and let the type of $\y$ be $\qtype$, the
optimal reproduction type for $P$ and the distortion metric $\dist$. Then
\[
P(\X\sim \y) 
\ge
\frac{(1-\dist/2)^2}{n}
\, 
P\Paren{B(\y,d)},
\]  
where, $\X\sim\y$ is as defined in text preceding this Lemma.

\Proof We adapt a so-called \emph{Cycle Lemma} in Dvoretzky and Motzkin~\cite{DM47} that has been rediscovered several
times~\cite{Ren07} in literature.

Consider sequences $\y_0$ and $\y_1$ corresponding to the zeros and ones of $\y$. We first look for sequences $\x_0$ and
$\x_1$ satisfying $d(\x_0,\y_0)\le \dist$ and $d(\x_1,\y_1)\le \dist$, and make a sequence $\x$ by replacing the zeros
of $\y$ with $\x_0$ and the ones of $\y$ with $\x_1$. Let $\cB$ be the set of all such sequences $\x$.

Suppose $(\x_0)$ and $(\x_1)$ are cyclic shifts of some valid $\x_0$ and $\x_1$ respectively. Then the cycle lemma
of~\cite{DM47} states that at least $(1-\dist/2)$ fraction of these cyclic shifts are $\sim\y_0$ and $\sim\y_1$
respectively---we call them \emph{good} shifts. Note that if we replace both $\y_0$ and $\y_1$ with good shifts of
$\x_0$ and $\x_1$ to obtain a sequence $\x$, then it follows that $\x\sim\y$. In addition, all sequences formed by
replacing the zeros and ones with (good or otherwise) shifts of $\x_0$ and $\x_1$ have the same type, and hence the same
probability under $P$. Thus
\[
P(\X\sim \y) \ge (1-\dist/2)^2 P(\cB).
\]
Furthermore, it is easy to verify that if the type of $\y$ is the optimal reproduction type, (remove and use only
previous equation---the next equation is unnecessary and never used)
\[
P(\cB) \ge \frac1n P(B(\y,\dist)).\eqed
\]
\eLemmap

\section{Codelet parsing}
\label{sec:algo}
At the core of the paper is the \emph{Codelet parsing} algorithm for lossy compression with a Hamming distortion
constraint.  When no distortion is allowed, the algorithm reduces to the lossless Lempel Ziv algorithm. Codelet Parsing
{\em sequentially} parses the source sequence into non-overlapping phrases, mapping each phrase to a \emph{codelet} in a
\emph{dictionary}. The dictionary in turn is updated.

At the block level, the codelet parsing algorithm maps a source sequence $x_1^n$ to a distorted sequence $y_1^n$, and
then encodes and transmits the latter without loss using a LZ78 encoder. We describe the algorithm with an example, full
details are available in~\cite{MS05}.

\bExample Consider the string $x_1^{13}=0110101101000$, which we will encode with allowable hamming distortion $\dist\le
1/2$. We initialize a codebook $\cC_0=\sets{0,1}$, call the members of the codebook as \emph{codelets}, and denote the
type of a string $v$ by $\tau (v)$.  At each step, we choose a codelet to represent a portion of the unparsed string,
such that the codelet is within distortion 1/2 from a matching length prefix of the unparsed string.

At step $t=1$, the unparsed string is 0110101101000. The codelet 0 has a prefix (0) within distortion 0, while the
codelet 1 does not match any prefix to within distortion 1/2. The first bit of $x_1^{13}$ is represented by the codelet
0, and the matching codelet 0 in $\cC_0$ is replaced by its one bit extensions, namely 00 and 01, to yield $\cC_1$.

Now $\cC_1=\sets{00,01,1}$, and the unparsed segment of the string is 110101101000. Note that codelet 1 has a prefix (1)
within distortion 0 while the codelet 01 has a prefix (11) within distortion 1/2. We have two choices: represent the
first bit of the unparsed segment with the codelet 1, or the first two bits of the unparsed segment with 01.

To decide, we build the set of matching codelets $\cM_1=\sets{01,1}$.  To each codelet $m\in\cM_1$, associate the prefix
$r$ of $x_1^{13}$ that will be parsed thus far if $m$ is chosen, and compute the metric
$\im(\tau(m),\tau(r),\dist)$. Therefore for $m=01$, the prefix $r$ of $x_1^{13}$ associated is 011 (0 from the first
round, and 11 from this round). The metric for the codelet 01 is then $\im(\tau(01),\tau(011),1/2)$.  Choose the codelet
with the minimum metric, and update the codebook by replacing the chosen codelet with its one bit extensions. Suppose
the chosen codelet is 01, $\cC_2=\sets{00,010,011,1}$, and the bits 11 are represented by 01 in this round. The unparsed
string for the next round is then 0101101000.  \eExample

As we saw in the second round above, there are usually multiple ways to parse the incoming source string and map it into
codewords. Indeed the crux of the algorithm is the answer to:
\begin{quote}
How do we select between multiple parsings?
\end{quote}
Interestingly, the most natural extension of Lempel Ziv algorithm to the lossy case---picking one of the longest codelet
among the matches---is proven suboptimal in~\cite{luczak:szpankowski}, in a specific LZ77 setting.

\section{Idealization of codelet parsing}
To understand the codelet parsing algorithm described above, we idealize the codelet parsing algorithm in order to
isolate the core phenomena underlying the algorithm, and to make it amenable to a simple analysis.

(remove, add universal section) For the sake of simplicity, and because we are only analyzing the \iid case in this
paper, we assume that the Idealized Codelet Parsing algorithm knows the underlying statistics of the data. Note that in
the \iid case, we learn the underlying statistics at the rate of $\cO(1/\sqrt{s})$, where $s$ is the length of the
string we have observed thus far, and hence at an exponentially faster rate than we would expect for any LZ type
algorithm.

\subsection*{Modifications}
\setcounter{paragraph}{0}
\paragraph{Known horizon}
First, we assume that the blocklength of the input string $\vecx$ is known in advance. Note that while this aids
analysis, it is not a stringent restriction. In practice, a modification of the \emph{doubling trick}~(\cite{CL},
Chapter 2.3) can be used to handle strings whose length is unknown, with asymptotically no degradation in
performance. For details, please see~\cite{SM11:all}.

Let $\y$ be a length-$L$ sequence with the optimal reproduction type,
and let 
\[
\pl=\inpd(\X\sim \y),
\]
where $\X$ is a sequence generated by $\inpd$.  Now let
\[
\Ml = L^2/\pl.
\]
Further, denote an input sequence ${\bf z}$ of length $\ell$ to be $\epsilon$-typical if $|h(p)+\log \inpd({\bf z})|\le
\ell \epsilon$, and let $T_\X^{\ell,\epsilon}$ be the set of all $\ell-$length $\epsilon-$typical sequences.

\paragraph{Updating the dictionary} 
The Idealized Codelet Parsing algorithm initializes the dictionary with all $2^\ell$ $\ell-$length sequences. Among
them, it first obtains a set $\cD_\ell$ of $M_\ell$ codelets of length $\ell$.  Then, every sequence in $\cD_\ell$ is
replaced with all its $2^\ell$ $\ell$-bit extensions, and among them $M_{2\ell}$ length-$2\ell$ codelets are chosen to
obtain $\cD_{2\ell}$. The algorithm proceeds by then updating the dictionary with longer codelets, forming in turn, the
sets $\cD_{k\ell}$ for increasing values of $k$.

\paragraph{Selecting codelets by partial matching} 
\label{s:rules}
To pick any codelet to represent a portion of the unparsed, input sequence, the algorithm finds the longest matching
codelet from the leaves of the dictionary tree.

Note that we can exploit because we map any codelet $\y$ to only sequences $\x$ such that $\x\sim\y$, finding the
longest match does \emph{not} require exhaustive search among the codelets with high probability. Following is an
algorithm that does the search among $L$-length codelets in $\cO(2^\ell L^2)$ operations with high probability.

Let $\x=x_1,x_2\ldots$ be the unparsed segment of the input.
\[
Z_\ell=\sets{\y \in \cD_\ell: \y \sim x_1^\ell }.
\]
be the partial matches at level $\ell$. Among all the descendents of $Z_\ell$ in $\cD_{2\ell}$, find all partial matches
for $x_1^{2\ell}$ to obtain $Z_{2\ell}$. The crucial point to observe is

\bProperty If
there exists $y_1^{2\ell}\in\cD_{2\ell}$ such that $y_1^{2\ell}\sim
x_1^{2\ell}$, then $y_1^{\ell}\sim x_1^{\ell}$, namely $y_1^\ell\in
Z_{\ell}$.  \eProperty 

Therefore $Z_{2\ell}$ contains all sequences in $\cD_{2\ell}$ that $\sim x_1^{2\ell}$. We would not have this property
if we simply obtained the sets $Z$ by picking codelets that satisfied the distortion constraint alone. Combined with the
Lemma~\ref{lm:ms} below that with high probability, $|Z_{k\ell}|$ grows polynomially rather than exponentially,
obtaining $Z_L$ for any $L$ can be done polynomially in $L$. In the low probability event that $Z_{k\ell}$ grows faster
than the Lemma bound, we simply give up.

\bLemma
For all $\delta$, with probability $\ge 1-\delta$, simultaneously for all $k$ 
\[
|Z_{k\ell}|\le \frac{(k\ell)^4}\delta.\eqed
\]
\eLemmap

\section{Optimality of Codelet Parsing}
We show that the Idealized Codelet Parsing algorithm is optimal.  Let $X^n=X_1\upto X_n$ be generated by a binary
memoryless source $\inpd$, with $\inpd(X_1=1)=p$. Let the \emph{target} average Hamming distortion constraint be
$\dist$. Let $Y^n$ be the distorted representation of $X^n$ output by the algorithm, and let $\mathcal{L}(Y^n)$ be the
number of bits required to describe $Y^n$. Then, \bTheorem
\label{thm:main}
For the Idealized Codelet Parsing algorithm,
\[
\frac1n \E\mathcal{L}(Y^n) \le \rd +\cO\Paren{\frac{\log\log n}{\log n}},
\]
and $\frac1n d(X^n,Y^n)\le \dist$ 
\eTheorem 
The expectation above is taken over all the choices made by the algorithm and over the input sequences.

\subsection*{Analysis of the cover}
We first establish that the codelets provide a good cover for the source phrases.

The algorithm chooses codelets of lengths $\ell, 2\ell$ and so on.  We will often refer to the length of codelets as
their \emph{depth}, since they are either internal nodes or leaves of the dictionary tree.  Let $\rY^L_i$ be the $i'th$
(in sequence) codelet chosen at depth $L$ of the dictionary tree. Note that the dictionary is itself random (dictated by
$\X$ and the random choices made while populating it), and we denote by $\cD_L$ the dictionary at depth $L$ once the
algorithm has processed a length $n$ sequence.  For any sequence $\X$ with length $L$, let $\Nl(\X)$ be the number of
codelets in $\cD_L$ that are within the distortion budget from $\X$. We will drop the argument of $\Nl$ when writing
expectations for simplicity. All expectations that follow are over $\X$ and $\cD$.

As mentioned before, too many matches is a sign of suboptimality. To quantify this, we compute $\E \Nl$ and $\E
\Nl^2$. Together, they provide a lower bound on the probability $\Nl>0$, namely the probability that $\X$ is covered by
some element of the dictionary at depth $L$.

Clearly $\E \Nl$ is easy to compute for any $L$ by linearity of expectation.  However $\E \Nl^2$ is somewhat trickier to
bound, but is well behaved. We show in Lemma~\ref{lm:nlsq} that when averaged over all possible codebooks, $\E \Nl^2$ is
lower than the corresponding expectation if we chose $\Ml$ codelets at random. From~\cite{}, random choice of codelets
leads to good covers with overwhelming probability.  We will therefore conclude that, the cover gets better as we parse
longer. Computation of $\E \Nl^2$ is somewhat involved, but the algebra is simplified for a Bernoulli $1/2$ source.

We first note that the codebook construction contains symmetries that we will need to exploit for Lemma~\ref{lm:nlsq}.

\bLemma
\label{lm:sym}
Let $\Nl=\binom{L}{Lq}$. For all $y\in T^L_q$, 
\[
\prob(y\in \cD_L)=\frac{\Ml}{\Nl}
\]
\Proof Suppose the length of $y$ be $L=k\ell$ and let $y'={y'}_1^{\ell}, {y'}_{\ell+1}^{2\ell}\upto
{y'}_{(k-1)\ell+1}^L$. 
Note that each ${y'}_{i\ell}^{(i+1)\ell}$ can be obtained from the corresponding subsequence $y_{i\ell+1}^{(i+1)\ell}$
by some permutation of bit locations of the later, since both bit sequences have the same type. Represent these
permutations by $\sigma_0 \upto \sigma_{k-1}$, and we write ${y'}_1^\ell=\sigma_0(y_1^\ell)$ as a shorthand. These
permutations are not unique, however we will fix one valid value for each of $\sigma_0\upto \sigma_{k-1}$.

Let $XC(y)$ be the set of length $n$ input sequences and the corresponding choices between multiple matches made by the
algorithm that induce $y\in \cD_L$. Corresponding to each input sequence $\x$ that could induce $y$, we represent the
choices as numbers, one for each phrase, indicating (in lexicographic order) which of the codelets that $\sim\x$ are
chosen. Thus,
\[
XC(y)= \sets{ (\x, \c): \text{ choices $\c$ on sequence $\x$ induce $y$ } }.
\]
Similarly for $XC(y')$.

To see that there is a bijection between $XC(y')$ and $XC(y)$, take an element $(\x,\c) \in X(y)$. From $\x$, we obtain
$\x'\in XC(y')$ by manipulating each phrase obtained in the parsing of $\x$. Suppose $z=z_1^{\ell}\ldots
z_{(m-1)\ell+1}^{m\ell}$ is a phrase obtained during the parsing of $\x$.  If $m\le k$ we replace $z$ with
\[
z'=\sigma_0(z_1^{\ell})\ldots\sigma_{m-1}(z_{(m-1)\ell}^{m\ell})
\] 
and if $m>k$ we replace $z$ with
\[
z'=\sigma_0(z_1^{\ell})\ldots\sigma_{k-1}(z_{(k-1)\ell}^{k\ell})z_{k\ell+1}^{(k+1)\ell}
\ldots z_{(m-1)\ell+1}^{m\ell}.
\]
Now to make choices among competing matches, instead of lexicographic ordering, we use the lexicographic ordering under
$\prod\sigma_i^{-1}(z_{(i-1)\ell}^{i\ell})$ (replace $\prod$ with concatenation symbol).  Now, note that if $(\x,\c)$
yielded $y$, $(\x',\c)$ will yield $y'$. Finally, since \iid probabilities of sequences do not change when their bit
locations are permuted, it follows that
\[
\prob(y\in \cD_L) = \prob(XC(y)) = \prob(XC(y')) = \prob(y'\in \cD_L).\eqed
\]
\eLemmap

\ignore{To dig a little deeper into the symmetries underlying codelet parsing, we observe that
\[
\E 1(\text{ $y$ added to $\cD_L$ in step $i$ })
\]
has the same value for all $y\in T^L_q$.  Denote the above by $q_i$.  Thus, from the first part of this Theorem,
\[
\sum_{i=1}^M q_i = \frac{\Ml}{N}.
\]
Let $F^{L}_q$ be the set of sequences of length $L$ and type $q$ that extend the codelets $\cD_{L-\ell}$ chosen at
length $L-\ell$. These sequences can potentially be chosen as codelets with length $L$. Now for any $l$ and any $y\in
F^{L}_q$,
\begin{equation}
\label{eq:prob}
\prob( Y_{l}=y |Y_1\upto Y_{l-1})
=
\sum_{\x\in B(y,d) } 
\inpd(\x)\frac1{\nm(\x)} 
1\Paren{\x\notin \cup_{i=1}^{l-1} B(Y_i,d)},
\end{equation}
where 
\[
\nm(\x) = | \sets{ y \in F^L_q: y\in B(\x,d) } |
\]
is the set of feasible codelets within distortion $d$ from $\x$.  Suppose we remove $Y_{1}$ from $\cD_L$. Namely,
consider the first $l-2$ codelets chosen to be $Y_2\upto Y_{l-1}$. From~\eqref{eq:prob} we therefore have
\[
\prob( Y_{l}=y |Y_1\upto Y_{l-1})\le \prob(Y_{l-1}=y | Y_2\upto Y_{l-1}).
\]
There is, of course, no sanctity to the index being 1---removing $Y_j$ instead yields
\bProperty
\label{prop:crucial}
\[
\prob( Y_{l}=y |Y_1\upto Y_{l-1})\le \prob(Y_{l-1}=y | Y_1\upto Y_{j-1}, Y_{j+1}\upto Y_{l-1}).
\]
\eProperty 
Therefore, it can also be shown that $q_i$ is decreasing in $i$. More importantly, it follows that}
\bLemma
Let $y_1,y_2\in T^L_q$ be identical in the first $r$ $\ell-$length segments. Then,
\[
\prob(\,y_1\text{ and } y_2 \in \cD_L \,)
\le
\frac{\Ml}{\Nl}\frac{\Ml}{\Mr\Nlmr}.\eqed
\]
\eLemmap

The next Lemma would easily follows from the linearity of expectation, but we provide a slightly more convoluted proof
using the above Lemma~\ref{lm:sym}. Let $N_{L,\cD}(\X)$ be the number of codelets that match $\X$ in the randomly chosen
codebook $\cD$. For the codelet parsing algorithm described above,

\bLemma
\label{lm:nl}
$\E N_{L,\cD} = \Ml \pl$.
\Proof Note that
\begin{align*}
\E N_{L,\cD} &= \sum_{\x} \inpd(\x)\sum_{y} 1(y\in\cD_L\text{ and } y\sim \x)=\sum_{y}\prob(y\in\cD_L) \sum_{\x\in B(y,\dist)} \inpd(\x) \\
&\aeq{(a)}\sum_{y} \frac{\Ml}{\Nl} \prob(B(y,d)) =\Ml \pl.
\end{align*}
where $(a)$ follows from Lemma~\ref{lm:sym}.
\eLemma

\bLemma
\label{lm:intx}
Let $\y_L$ and $\tilde{\y}_L$ be two sequences with type $q$. Let $y_\ell$ and $\tilde{y}_\ell$ be
two sequences with type $q$ and length $\ell$. Then,
\[
P\Paren{ B(\y_L y_\ell, d) \cap B(\tilde{\y}_L\tilde{\y}_{\ell},d) }
\le
P\Paren{ B(\y_L, d) \cap B(\tilde{\y}_L,d) }
\Paren{\frac{\plpl}{\pl}}^2.\eqed
\]
\eLemmap

\bLemma
\label{lm:nlsq}
Let $N_{L,\cD}(\X)$ be the number of codelets of length $L$ that match $\X$ in codebook
$\cD_L$, and let $N_{L,\cD}(\X)$ be the number of codelets that match $\X$ in a codebook $\cD$. Then
\[
\E N^2_{L+\ell,\cD} 
\le
(\E N^2_{L,\cD} +\E N_{L,\cD})\Paren{\frac{ \Mlpl\plpl}{\Ml\pl}}^2 
\]
where $\pl = \prob(B(y,d))$ for any $y\in T^L_q$. 
\ignore{\Proof First observe that
\begin{align*}
&E N_{L+\ell,\cD}^2
=\sum_{\x} \inpd(\x) N_{L+\ell}(\x)^2\\
&=
\E_\cD
\sum_{\x} 
\inpd(\x) 
\Paren{\sum_{y\in T^L_q} 1(y\in \cD_L) 1(\x\sim y)}
\Paren{\sum_{y'\in T^L_q} 1(y'\in \cD_L) 1(\x\sim y')}\\
&=
\sum_{\substack{y,y'\in T^L_q\\y\ne y'}} 
\E_\cD 1(y'\in \cD_L \text{ and } y\in \cD_L) 
\sum_{\x}
\inpd(\x) 
1(\x\sim y)
1(\x\sim y')
+
\sum_{\substack{y''\in T^L_q}} 
\E_\cD 1(y''\in \cD_L )
\sum_{\x}
\inpd(\x) 
1(\x\sim y'')\\
&=
\sum_{\x}
\inpd(\x) 
\sum_{\substack{y,y'\in T^L_q\\y\ne y'}} 
\prob(y'\in \cD_L \text{ and } y\in \cD_L) 
1(\x\sim y)1(\x\sim y')
+
\Ml \pl\\
&\ale{(a)}
\frac{2\Ml^2}{\Nl^2}
\sum_{\x}
\inpd(\x) 
\sum_{\substack{y,y'\in T^L_q\\y\ne y'}} 
1(\x\sim y)1(\x\sim y')
+
\Ml \pl\\
&\ale{(b)}
\frac{2\Ml^2}{\Nl^2}
(\Nl \pl)^2
\sum_{\x}
\inpd(\x) 
+
\Ml \pl\\
&\le
2\Ml^2\pl^2
+
\Ml \pl.\tag*{$\Box$}
\end{align*}}
\eLemma

For comparison let us consider the expected value of $\E N_{L,\cD}^2$ for random codebook constructions of length $L$.
Here we use codebooks $\cC_L$ populated with sequences of type $q$ as follows. Generate independent sequences of length
$L$, with the $L$-length sequence generated in step $(i)$ being $\X^{(i)}$. Each $\X^{(i)}$ is in turn obtained by
generating $L$ bits \iid Bernoulli $(1/2)$. Initialize $\cC^{(0)}_L=\phi$.  At every step $i$, update
$\cC^{(i)}=\cC^{(i-1)}_L\cup\sets{ y}$, where $y$ is a randomly chosen length $L$ sequence of type $q$ such that
$\X^{(i)}\in B(y,\dist)$. Stop after the $i=\Ml'$th codelet is chosen, and let $\cC_L=\cC^{(\Ml)}_L$.  For such a random
codebook construction, it is easy to see that
\[
\prob(y'\in \cC_L \text{ and } y\in \cC_L) 
=
\frac{\Ml (\Ml-1)}{\Nl(\Nl-1)}.
\]

The above Lemmas imply
\bCorollary
\label{corr:main}
$P(N_{L+\ell,\cD} > 0)  \ge \frac{P(N_{L,\cD}>0)}{P(N_{L,\cD}>0)+1/(\Ml \pl)}$
\Proof Cauchy Schwartz Inequality.
\eCorollary

The next cog in the proof is the observation that there cannot be too many ``short'' phrases in the lossy
representation.  
\bLemma
\label{lm:number}
For $n$ sufficiently large, the number of nodes in the dictionary with length shorter than $\frac{\log
  n-7\ell}{R(d)}$ is $\le \frac{n}{(\log n)^2}$.  
\eLemma

The details of the reminder of the proof is omitted, but follows the following line of arguments standard in LZ analysis
literature. (Complete below)

The section populated by short phrases contributes at most redundancy $1/(\log n)$.  Unrolling
Corollary~\ref{corr:main}, with high probability we find that some element of the dictionary matches an incoming
phrase. Describing such phrases takes at most $\log n$ bits, and such phrases by Lemma~\ref{lm:number} have length
$\ge {\log n-7\log\log n}/R(D)$, yielding a per symbol encoding rate of $R(D)+\cO\Paren{frac{\log\log n}{\log n}}$.
With a small probability, no element of the dictionary matches an incoming phrase---forcing us to describe such
phrases bit for bit, adding another $\cO\Paren{frac{\log\log n}{\log n}}$ to the coding rate.

(Complete above)

\section{Acknowledgments}
We thank L. Lastras-Monta\~{n}o for helpful discussions and constructive suggestions, as well as D. Baron, Y. Kochman,
J. {\O}stergaard, and G. Wornell for helpful discussions.  

\bibliographystyle{unsrt}
\bibliography{univcod,modha,mjwain_super}

\begin{thebibliography}{10}

\bibitem{Kac47}
M.~Kac.
\newblock On the notion of recurrence in discrete stochastic processes.
\newblock {\em Bulletin of the American Math Society}, 53:1002--1010, Oct 1947.

\bibitem{Ziv08}
J.~Ziv.
\newblock On finite memory universal data compression and classification of
  individual sequences.
\newblock {\em IEEE Transactions on Information Theory}, 54(4):1626--1636,
  2008.

\bibitem{WOSVW05}
T.~Weissman, E.~Ordentlich, G.~Seroussi, S.~Verdu, and M.~Weinberger.
\newblock Universal discrete denoising: known channel.
\newblock {\em IEEE Transactions on Information Theory}, 51(1):5--28, 2005.
\newblock See also HP Labs Tech Report HPL-2003-29, Feb 2003.

\bibitem{LZ76}
A.~Lempel and J.~Ziv.
\newblock On the complexity of finite sequences.
\newblock {\em IEEE Transactions on Information Theory}, 22:75--81, 1976.

\bibitem{KS87}
F.~Kaspar and H.~Schuster.
\newblock Easily calculable measure for the complexity of spatiotemporal
  patterns.
\newblock {\em Phys. Rev. A}, 36(2):842--848, Jul 1987.

\bibitem{SM11:all}
N.~Santhanam and D.~Modha.
\newblock Lossy lempel-ziv like compression algorithms for memoryless sources.
\newblock In {\em Allerton Conference on Computing, Communication and Control},
  September 2011.

\bibitem{berger:gibson}
T.~Berger and J.~D. Gibson.
\newblock Lossy source coding.
\newblock {\em IEEE {T}rans. {I}nform. {T}heory}, 44(6):2693--2723, 1998.

\bibitem{kieffer:93}
J.~C. Kieffer.
\newblock A survey of the theory of source coding with a fidelity criterion.
\newblock {\em IEEE {T}rans. {I}nform. {T}heory}, 39(5):1473--1490, 1993.

\bibitem{gray:neuhoff}
Robert~M. Gray and David~L. Neuhoff.
\newblock Quantization.
\newblock {\em IEEE {T}rans. {I}nform. {T}heory}, 44(6):2325--2383, 1998.

\bibitem{atallah:01}
M.~J. Atallah, F.~Chyzak, and P.~Dumas.
\newblock A randomized algorithm for approximate string matching.
\newblock {\em Algorithmica}, 29:468--486, 2001.

\bibitem{navarro}
G.~Navarro.
\newblock A guide to approximate string matching.
\newblock {\em ACM Computing Surveys}, 33(1):31--88, 2001.

\bibitem{cheung:wei}
K.~Cheung and V.~K. Wei.
\newblock A locally adaptive source coding scheme.
\newblock In {\em Bilkent Conf. on New Trends in Comm., Cont., and Signal
  Proc.}, pages 1473--1482, 1990.

\bibitem{yang:kieffer:98}
En-Hui Yang and J.~C. Kieffer.
\newblock On the performance of data compression algorithms based upon string
  matching.
\newblock {\em IEEE {T}rans. {I}nform. {T}heory}, 44:47--65, 1998.

\bibitem{zhang:wei:96}
Z.~Zhang and V.~K. Wei.
\newblock An on-line universal lossy data compression algorithm via continuous
  codebook refinement--part i: Basic results.
\newblock {\em IEEE {T}rans. {I}nform. {T}heory}, 42(3):803--821, 1996.

\bibitem{morita:kobayashi}
H.~Morita and K.~Kobayashi.
\newblock An extension of {LZW} coding algorithm to source coding subject to a
  fidelity criterion.
\newblock In {\em Proc.\ 4th Joint Swedish-Soviet Int. Workshop on Information
  Theory, Gotland, Sweden}, pages 105--109, 1989.

\bibitem{cornel:storer:93}
C.~Constantinescu and J.~A. Storer.
\newblock On-line adaptive vector quantization with variable size codebook
  entries.
\newblock In {\em Data Compression Conf.}, pages 32--41, 1993.

\bibitem{cornel:94}
C.~Constantinescu and J.~A. Storer.
\newblock Improved techniques for single-pass vector quantization.
\newblock {\em Proceedings of the IEEE}, 82(6):933--939, 1994.

\bibitem{storer:88}
J.~A. Storer.
\newblock {\em Data Compression: Methods and Theory}.
\newblock Computer Science Press, Rockville, Maryland, 1988.

\bibitem{steinberg:gutman}
Y.~Steinberg and M.~Gutman.
\newblock An algorithm for source coding subject to a fidelity criterion, based
  on string matching.
\newblock {\em IEEE {T}rans. {I}nform. {T}heory}, 39(3):877--886, 1993.

\bibitem{luczak:szpankowski}
T.~Luczak and W.~Szpankowski.
\newblock A suboptimal lossy data compression based on approximate pattern
  matching.
\newblock {\em IEEE {T}rans. {I}nform. {T}heory}, 43:1439--1451, 1997.

\bibitem{kontoyiannis}
I.~Kontoyiannis.
\newblock An implementable lossy version of the {L}empel-{Z}iv algorithm--part
  i: Optimality for memoeyless sources.
\newblock {\em IEEE {T}rans. {I}nform. {T}heory}, 45(7):2293--2305, 1999.

\bibitem{genin}
M.~Atallah, Y.~Genin, and W.~Szpankowski.
\newblock Pattern matching image compression: Algorithmic and empirical
  results.
\newblock In {\em Proc. Int. Conf. Image Processing, Lausanne, Switzerland},
  volume~II, pages 349--352, 1996.

\bibitem{kieffer:yang:seq}
J.~C. Kieffer and E.-H. Yang.
\newblock Sequential codes, lossless compression of individual sequences, and
  {K}olmogorov complexity.
\newblock {\em IEEE {T}rans. {I}nform. {T}heory}, 42(1):29--39, 1996.

\bibitem{alzina}
M.~Alzina, W.~Szpankowski, and A.~Grama.
\newblock {2D}-pattern matching image and video compression.
\newblock In {\em Data Compression Conf.}, pages 424--433, 1999.

\bibitem{zamir:rose:haifa}
R.~Zamir and K.~Rose.
\newblock Towards lossy {L}empel-{Z}iv: Natural type selection.
\newblock In {\em Proc. Inform. Theory Workshop, Haifa, Israel}, page~58, 1996.

\bibitem{zamir:rose}
R.~Zamir and K.~Rose.
\newblock Natural type selection in adaptive lossy compression.
\newblock {\em IEEE {T}rans. {I}nform. {T}heory}, 47(1):99--111, 2001.

\bibitem{kochman:zamir}
Y.~Kochman and R.~Zamir.
\newblock Adaptive parametric vector quantization by natural type selection.
\newblock In {\em Data Compression Conference}, pages 392--401, 2002.

\bibitem{yang:kieffer:96}
En-Hui Yang and J.~C. Kieffer.
\newblock Simple universal lossy data compression schemes derived from the
  {L}empel-{Z}iv algorithm.
\newblock {\em IEEE {T}rans. {I}nform. {T}heory}, 42(1):239--245, 1996.

\bibitem{yang:zhang:mar99}
E.-H. Yang and Z.~Zhang.
\newblock Variable-rate trellis source encoding.
\newblock {\em IEEE {T}rans. {I}nform. {T}heory}, 45(2):586--608, 1999.

\bibitem{MatYam03}
Y.~Matsunaga and H.~Yamamoto.
\newblock A coding theorem for lossy data compression by {LDPC} codes.
\newblock {\em IEEE Trans. Info. Theory}, 49:2225--2229, 2003.

\bibitem{WaiMan05}
M.~J. Wainwright and E.~Maneva.
\newblock Lossy source coding by message-passing and decimation over
  generalized codewords of {LDGM} codes.
\newblock In {\em International Symposium on Information Theory}, Adelaide,
  Australia, September 2005.
\newblock Available at arxiv:cs.IT/0508068.

\bibitem{MarWai06a}
E.~Martinian and M.~J. Wainwright.
\newblock Low density codes achieve the rate-distortion bound.
\newblock In {\em Data Compression Conference}, volume~1, pages 153--162, March
  2006.
\newblock Available at arxiv:cs.IT/061123.

\bibitem{DimWaiRam07}
A.~G. Dimakis, M.~J. Wainwright, and K.~Ramchandran.
\newblock Lower bounds on the rate-distortion function of {LDGM} codes.
\newblock In {\em Information Theory Workshop}, September 2007.

\bibitem{JMW10}
S.~Jalal, A.~Montanari, and T.~Weissman.
\newblock Lossy compression of discrete sources via viterbi algorithm, 2010.
\newblock arXiv:1011.3761v2 [cs.IT] 21 Nov 2010.

\bibitem{modha:dcc:03}
D.~S. Modha.
\newblock Codelet {P}arsing: Quadratic-time, sequential, adaptive algorithms
  for lossy compression.
\newblock In {\em Proc. DCC, Snowbird, UT}, March 24--27, 2003.

\bibitem{modha:lossy:RJ2}
D.~S. Modha.
\newblock The art of making mistakes: A quadratic-time, sequential, adaptive
  algorithm for lossy compression.
\newblock Technical Report RJ 10286, IBM Almaden Research Center, San Jose, CA,
  February 19, 2003.

\bibitem{MS05}
D.~Modha and N.P. Santhanam.
\newblock Making the correct mistakes.
\newblock In {\em Proceedings of the Data Compression Conference}, 2006.

\bibitem{DM47}
A.~Dvoretzky and Th. Motzkin.
\newblock A problem of arrangements.
\newblock {\em Duke Mathematics Journal}, 14:305--–313, 1947.

\bibitem{Ren07}
M.~Renault.
\newblock Four proofs of the ballot theorem.
\newblock {\em Mathematics magazine}, 80(5), December 2007.

\bibitem{CL}
N.~Cesa Bianchi and G.~Lugosi.
\newblock {\em {Prediction, Learning and Games}}.
\newblock Cambridge University Press, 2006.

\end{thebibliography}
\end{document}